\documentclass[12pt]{article}

\usepackage[body={17.5cm, 21cm},right=2cm]{geometry}

\usepackage{color}
\usepackage{graphicx}
\usepackage{epsf}
\usepackage{graphicx,epsfig}
\usepackage{amsmath}
\usepackage{amssymb}
\usepackage{epsfig}
\usepackage{cite}
\usepackage{color,colordvi}
\newcommand{\be}{\begin{eqnarray}}
\newcommand{\ee}{\end{eqnarray}}
\newcommand{\bi}{\begin{itemize}}
\newcommand{\ei}{\end{itemize}}

\newcommand{\bx}{{\vec{x}}}

\let\latexcirc=\circ
\newcommand{\ccirc}{\mathbin{\mathchoice
  {\xcirc\scriptstyle}
  {\xcirc\scriptstyle}
  {\xcirc\scriptscriptstyle}
  {\xcirc\scriptscriptstyle}
}}
\newcommand{\xcirc}[1]{\vcenter{\hbox{$#1\latexcirc$}}}
\let\circ\ccirc

\newcounter{hran}

\def\MSbar{\relax\ifmmode\overline{\rm MS}\else{$\overline{\rm MS}${ }}\fi}

\def\d{\rm d}

\def\d{{\rm d}}

\def\vq{\vec{q}}
\def\vx{\vec{x}}
\def\vv{\vec{v}}

\def\bx{{\vec{x}}}
\def\bk{{\vec{k}}}
\def\bnabla{{\vec{\nabla}}}
\def\bv{{\vec{v}}}
\def\t{\tau}

\def\r{\rho}
\def\t{\tau}

 \def\vx{\vec{ x}}
  \def\vs{\vec{ s}} 
\def\vk{\vec{k}}
\def\vn{\vec{n}}
\def\vq{\vec{q}}
\def\vy{\vec{y}}
\def\hx{\hat{x}}
\def\r{R_L}

\def\simlt{\stackrel{<}{{}_\sim}}
\def\simgt{\stackrel{>}{{}_\sim}}

\newcommand{\bmmmm}[4]{\langle \delta(#1) \delta(#2) \delta(#3) \delta(#4)\rangle'}
\newcommand{\bggg}[3]{\langle \delta_{\rm g}(#1) \delta_{\rm g}(#2) \delta_{\rm g}(#3)\rangle'}
\newcommand{\bmmm}[3]{\langle \delta(#1) \delta(#2) \delta(#3)\rangle'}
\newcommand{\pgg}[2]{\langle \delta_{\rm g}(#1) \delta_{\rm g}(#2)\rangle'}
\newcommand{\pmm}[2]{\langle \delta(#1)\delta(#2) \rangle'}

\newcommand{\pmml}[2]{\langle \delta^{(1)}(#1)\delta^{(1)}(#2) \rangle'}
\newcommand{\pmgl}[2]{\langle \delta^{(1)}(#1) \delta_{\rm g}^{(1)}(#2) \rangle'}
\newcommand{\pggl}[2]{\langle \delta_{\rm g}^{(1)}(#1) \delta_{\rm g}^{(1)}(#2) \rangle'}

\newcommand{\bijk}[1]{B^{#1}_{{\rm g}, q\rightarrow0}}

\def\sq{{q\rightarrow 0}}

\numberwithin{equation}{section}
\begin{document}
\thispagestyle{empty}
\vspace{5mm}
\vspace{0.5cm}
\begin{center}

\def\thefootnote{\fnsymbol{footnote}}

{\Large \bf 
Consequences of Symmetries and Consistency Relations \\
in  the Large-Scale Structure of the Universe\\
for Non-local bias and Modified Gravity\\
\vspace{0.25cm}	
}
\vspace{1.5cm}
{\large  
A. Kehagias$^{a,b}$, J.Nore\~{n}a$^{b}$, H. Perrier$^{b}$ and A. Riotto$^{b}$
}
\\[0.5cm]

\vspace{.3cm}
{\normalsize {\it  $^{a}$ Physics Division, National Technical University of Athens, \\15780 Zografou Campus, Athens, Greece}}\\

\vspace{.3cm}
{\normalsize { \it $^{b}$ Department of Theoretical Physics and Center for Astroparticle Physics (CAP)\\ 24 quai E. Ansermet, CH-1211 Geneva 4, Switzerland}}\\

\vspace{.3cm}


\end{center}

\vspace{3cm}

\hrule \vspace{0.3cm}
{\small  \noindent \textbf{Abstract} \\[0.3cm]
\noindent 
Consistency relations involving the soft limit of the $(n + 1)$-correlator functions of dark matter and galaxy overdensities can be obtained, both in real and redshift space,  thanks to the  symmetries enjoyed by the Newtonian equations of motion describing the 
 dark matter and galaxy fluids coupled through  gravity. We study the implications of such symmetries for the theory of galaxy bias
 and for the theories of modified gravity. We find that the invariance of the fluid equations under a coordinate transformation that induces a long-wavelength velocity constrain the bias to depend only on a set of invariants, while the symmetry of such equations under 
Lifshitz
 scalings in the case of matter domination allows one to compute the time-dependence of the coefficients in the bias expansion. We also find that in theories of modified gravity which violate the equivalence principle induce a violation of the consistency relation which may be a signature for their observation. Thus, given adiabatic Gaussian initial conditions, the observation of a deviation from the consistency relation for galaxies would signal a break-down of the so-called non-local Eulerian bias model or the violation of the equivalence principle in the underlying theory of gravity.

\vspace{0.5cm}  \hrule
\vskip 1cm

\def\thefootnote{\arabic{footnote}}
\setcounter{footnote}{0}


\baselineskip= 18pt

\newpage 

\section{Introduction}\pagenumbering{arabic}
Symmetries play a crucial role    in  understanding   the properties of  a physical system and they have turned out to be quite useful in characterizing the  cosmological perturbations  generated during a de Sitter  stage \cite{lrreview}.  Since the  de Sitter isometry group SO(1,4) acts  like  conformal group  on $\mathbb{R}^3$ when the fluctuations are on 
super-Hubble scales, the correlators of scalar fields, which are not the inflaton,  are constrained by conformal invariance   
\cite{antoniadis,maldacena1,creminelli1,us1,us2}.  
The  fact that the de Sitter isometry group acts as conformal group on the three-dimensional Euclidean space on super-Hubble scales can be also used to predict  the shape of the correlators involving the inflaton and  vector fields \cite{vec}.
Furthermore, if the  inflationary perturbations are generated in single-field models of inflation,   there exist conformal consistency relations among the inflationary correlators \cite{creminelli2,hui,baumann1,baumann2,mcfadden,mata,hui2,ber}.   

Consistency relations involving the soft limit of the $(n + 1)$-correlator functions of matter and galaxy overdensities have  also been  proposed   by investigating  the symmetries enjoyed by the Newtonian equations of motion of the non-relativistic dark matter and galaxy  fluids coupled to gravity
\cite{KR,ppls}. These consistency relations have been recently generalized to the relativistic limit \cite{jorge} (see also \cite{acc}), based on the observation that a long mode, in single-field models of inflation, reduces to a diffeomorphism since its freezing during inflation all the way until the late universe, even when the long mode is inside the horizon (but out of the sound horizon).

The large-scale consistency relations have the virtue of being true also for the galaxy overdensities, independently of the bias
between galaxy and dark matter. As such, they may serve as a guidance in building up a bias theory. Indeed, we will argue that the non-local Eulerian bias model can be seen as being built of quantities which are invariant under the symmetries enjoyed by the Newtonian fluid equations. Furthermore, they might be useful
in testing theories of modified gravity where    extra degrees of freedom  appear mediating extra long-range forces (other than the gravitational one) and possibly  leading to a violation of the Equivalence Principle (EP)  in the late universe and therefore to a violation of the consistency relation. In fact, assuming adiabatic Gaussian initial conditions, an observed violation of the consistency relations would either indicate a break-down of the non-local Eulerian bias model (and also the presence of terms in the effective fluid equations for galaxies that break the aforementioned symmetries), or a violation of the EP in the underlying theory of gravity.


It is in the spirit of exploring these topics that in this paper we aim to investigate what the large-scale consistency relations may tell us about  the galaxy bias 
and how they can be used to scrutinize modified gravity theories. In particular, we will show that the symmetries leading to the
consistency relations allow the presence of what is commonly dubbed non-local bias, that is a relation between the galaxy 
and the dark matter overdensities which is not a simple function of the local dark matter abundance. We will identify a series of invariants
(with respect to the symmetries) which should appear in the galaxy bias expansion, precisely because they are
allowed by the symmetries of the problem. Furthermore, we will investigate under which  conditions the consistency relations
are valid in the case in which a modification of gravity is attained far in the infrared on cosmological scales.

The paper is organized as follows. In section 2 we discuss the symmetries of the non-relativistic fluid equations for both 
dark matter and galaxies and we derive galaxy consistency relations for the $n$-point correlators of short wavelength modes in the 
background of a long wavelength mode perturbation. In section 3 we provide the invariants under the symmetries of the galaxy and 
dark matter fluids and we discuss their implication for the non-local bias. We also check that the galaxy consistency relation holds at tree- and one-loop level in the bias model. In section 4 we show how to extend the galaxy consistency relations to redshift space where actual experiments are made.  In section 5 we discuss the consequences of 
the symmetries for the  theories of modified gravity and how such modifications are imprinted in the $(n+1)$-point correlators in 
the squeezed limit.
Finally, section 6 presents our conclusions. 

\section{Symmetries and consistency relation of galaxy correlation functions in real space}

\noindent
Galaxies (or more precisely, some population thereof),  once formed, obey the following equations on sub-Hubble scales
 \be
 &&\frac{\partial \delta_{\rm g}(\bx,\t)}{\partial \t}+\bnabla\cdot[(1+\delta_{\rm g}(\bx,\t))\bv_{\rm g}(\bx,\t)]=0\label{fg1},\\
 && \frac{\partial \bv_{\rm g}(\bx,\t)}{\partial \t}+{\cal{H}}(\t)\bv_{\rm g}(\vx,\t)
 +[\bv_{\rm g}(\vx,\t)\cdot \bnabla]\bv_{\rm g}(\bx,\t)=\ -\bnabla\Phi(\bx,\t),\label{fg2}\\
 &&
 \nabla^2\Phi(\bx,\t)=\frac{3}{2} \Omega_{\rm m}{\cal{H}}^2(\tau)\delta(\bx,\tau), \label{fg3}
 \ee
where we have  denoted by $\bx$  the comoving spatial coordinates, 
$\tau=\int \d t/a$  the conformal time, $a$ the scale factor
in the FRW metric and  
${\cal{H}}=\d\ln a/\d\t$ is the conformal expansion rate.  In addition, 
 $\delta(\bx,\t)=(\rho(\bx,\t)/\overline{\rho}-1)$ is the overdensity over the mean matter density density $\overline{\rho}$, 
 $\delta_{\rm g}(\bx,\t)$ and $\bv_{\rm g}(\bx,\t)$ are the galaxy overdensity and peculiar velocity,
and $\Phi(\bx,\t)$ is the gravitational 
potential due to density 
fluctuations. Finally $
\Omega_{\rm m}=8\pi G\bar \rho a^2/3{\cal{H}}^2$
is the density parameter. Eq. (\ref{fg1}) assumes number conservation \cite{fry}. Eventually, one would like to go beyond the treatment
presented here in order to account for  phenomena like  formation and merging, which could be done for example by adding a source term to the right hand side of Eq. \eqref{fg1}. 

Dark matter is described by  a similar set of  non-relativistic fluid equations in 
the presence of gravity
 \be
 &&\frac{\partial \delta(\bx,\t)}{\partial \t}+
 \bnabla\cdot[(1+\delta(\bx,\t))\bv(\bx,\t)]=0\label{fl1},\\
 && \frac{\partial \bv(\bx,\t)}{\partial \t}+{\cal{H}}(\t)\bv(\vx,\t)
 +[\bv(\vx,\t)\cdot \bnabla]\bv(\bx,\t)=\ -\bnabla\Phi(\bx,\t),\label{fl2}\\
 &&
 \nabla^2\Phi(\bx,\t)=\frac{3}{2} \Omega_{\rm m}{\cal{H}}^2(\tau)\delta(\bx,\tau), \label{fl3}
 \ee
 Following Ref. \cite{KR}, one can show that in $\Lambda$CDM cosmology the set of equations (\ref{fg1}-\ref{fg3}) and 
 (\ref{fl1}-\ref{fl3}) is invariant under the 
 transformations (for a generic vector ${\vec{n}}(T)$)
 \be
 \t'=\t, ~~~\bx'=\bx+{\vec{n}}(T),
 \label{gen}
 \ee
  where
   \be
   T(\tau)=\frac{1}{a(\t)}\int^\t \d\eta \,a(\eta),
   \ee
provided that one transforms the fields as follows
\begin{align}
\delta_{\rm g}'(\bx,\t)&=\delta_{\rm g}(\bx',\t'),\label{gg1}\\
\bv_{\rm g}'(\bx,\t)&=\bv_{\rm g}(\bx',\t')-\dot{\vec{n}}(T),\label{gg2}\\
\delta'(\bx,\t)&=\delta(\bx',\t'),\label{dg1}\\
 \bv'(\bx,\t)&=\bv(\bx',\t')-\dot{\vec{n}}(T), \label{vg1}\\
\Phi'(\bx,\t)&=\Phi(\bx',\t')-\Big{(}{\cal{H}}\dot{\vec{n}}(T)+\ddot{\vec{n}}(T)\Big{)}\cdot \bx.\label{gg3}
\end{align}
This is true even if the we do not  set $\bv_{\rm g}(\bx,\t)=\bv(\bx,\t)$,  that is if we do not assume that   the galaxy peculiar velocity is unbiased. Note that if one adds a source term to the right hand side of Eq. \eqref{fg1} to account for the change of the number density of galaxies in time, and such a source term depends only on quantities which transform as scalars, the equations of motion are still invariant  under these transformations.

Consider the $n$-point correlation function of short modes of the density contrast. The symmetries of the Newtonian fluid equations imply, for instance, that 
\be
\Big<\delta_{\rm g}'(\vx_1)\cdots \delta_{\rm g}'(\vx_n)\Big>=\Big<\delta_{\rm g}(\vx_1)\cdots \delta_{\rm g}(\vx_n)\Big>=\Big<\delta_{\rm g}(\vx'_1)\cdots \delta_{\rm g}(\vx'_n)\Big>.
\ee
The points are supposed to be contained in a sphere of radius  $R$ much smaller than the long wavelength mode of size $\sim 1/q$ and centered at the origin of the coordinates.  The non-relativistic equations of motion are invariant under the generic transformation $\t\rightarrow \t$ and  $\bx\rightarrow \bx+{\vec{n}}(T(\tau))$. This means that we can generate a long wavelength mode for the dark matter  velocity perturbation $\bv_L(\t,\vec{0})$ just by choosing properly
the vector ${\vec{n}}(\tau)$
\be
{\vec{n}}(\tau)=-\int^\tau\d\eta\, \bv_L(\eta, \vec{0}) +{\cal O}(qR v_L^2).
\label{long}
\ee
In other words, the correlator of the short wavelength modes in the background of the long wavelength mode perturbation should satisfy the relation \cite{KR}
\be
\Big<\delta_{\rm g}(\t_1,\bx_1)\delta_{\rm g}(\t_2,\bx_2)\cdots\delta_{\rm g}(\t_n,\bx_n)\Big>_{v_L}=
\Big<\delta_{\rm g}(\t_1',\bx_1')\delta_{\rm g}(\t_2',\bx_2')\cdots\delta_{\rm g}(\t_n',\bx_n')\Big>.
\ee
This is nothing else that the statement that the effect of a physical long wavelength galaxy velocity perturbation  onto the short modes
should be  indistinguishable from the long wavelength mode velocity generated by the transformation with $\delta x^i={n}^i(\tau)$.
In momentum space one therefore obtains
\be
\Big< \delta_{\rm g}(\vq,\t)\delta_{\rm g}({\vk_1},\t_1)\cdots\delta_{\rm g}(\vk_n,\t_n)\Big>_{q\to 0}= 
\Big< \delta_{\rm g}(\vq,\t)
\Big<\delta_{\rm g}(\vk_1,\t_1)
\cdots\delta_{\rm g}(\vk_n,\t_n)\Big>_{v_L}\Big>. 
\ee
The variation of the  $n$-point correlator under the infinitesimal transformation  is given by
\begin{align}
\delta_n \Big<\delta_{\rm g}(\t_1,\bx_1)\cdots\delta_{\rm g}(\t_n,\bx_n)\Big>&=
\int \frac{\d^3\vk_1}{(2\pi)^3}\cdots \frac{\d^3\vk_n}{(2\pi)^3} \Big<\delta_{\rm g}(\vk_1,\t_1)
\cdots\delta_{\rm g}(\vk_n,\t_n)\Big>\nonumber \\
&\times \sum_{a=1}^n\delta x^i_a (i k_a^i) e^{i(\vk_1\cdot \vx_1+\cdots\vk_n\cdot \vx_n)}\nonumber \\
&=\int \frac{\d^3\vk_1}{(2\pi)^3}\cdots \frac{\d^3\vk_n}{(2\pi)^3} \Big<\delta_{\rm g}(\vk_1,\t_1)
\cdots\delta_{\rm g}(\vk_n,\t_n)\Big>\nonumber \\
&\times \sum_{a=1}^n n^i(\tau_a) (i k_a^i) e^{i(\vk_1\cdot \vx_1+\cdots\vk_n\cdot \vx_n)}.
\end{align}
Then we find that 
\begin{eqnarray}
\Big< \delta_{\rm g}(\vq,\t)\delta_{\rm g}(\vk_1,\t_1)\cdots\delta_{\rm g}(\vk_n,\t_n)\Big>_{q\to 0}&=&
\Big< \delta_{\rm g}(\vq,\t)\Big<\delta_{\rm g}(\vk_1,\t_1)\cdots\delta_{\rm g}(\vk_n,\t_n)\Big>_{v_L}\Big> \nonumber \\
&=&i\sum_{a=1}^n\Big< \delta_{\rm g}(\vq,\t) n^i(\tau_a)\Big>  k_a^i \Big<\delta_{\rm g}(\vk_1,\t_1)
\cdots\delta_{\rm g}(\vk_n,\t_n)\Big>.
\end{eqnarray}
In    a $\Lambda$CDM model we have
\begin{align}
\int^\tau\d\eta\, \bv_L({\vq},\eta)&=
i\frac{q^i}{q^2}\int^\tau\d\eta\,{\cal H}\,
\frac{1}{{\cal H}}\frac{\d\ln D(\eta)}{\d\eta}\,\frac{D(\eta)}{D(\eta_{\rm in})}\delta_L(\vq,\eta_{\rm in})
=i\frac{\vq}{q^2}\delta_L(\vq,\tau),
\end{align}
where $D(\t)$ is the linear growth factor and $\delta$ is the dark matter overdensity. We thus obtain the consistency relation



\be
\fbox{$\displaystyle
\Big< \delta_{\rm g}(\vq,\t)\delta_{\rm g}(\vk_1,\t_1)\cdots\delta_{\rm g}(\vk_n,\t_n)\Big>'_{q\to 0}
= -\Big< \delta_{\rm g}^L(\vq,\t) \delta_L(\vq,\t)\Big>'\sum_{a=1}^n \frac{D(\t_a)}{D(\tau)} \frac{{\vec q} \cdot \bk_a}{q^2}\Big<\delta_{\rm g}(\vk_1,\t_1)
\cdots\delta_{\rm g}(\vk_n,\t_n)\Big>'$},\nonumber\\
\label{deltadelta}\nonumber\\
\ee
where the primes indicate that one should remove the Dirac delta's coming from the momentum conservation. 
Notice that, if  the correlators are computed all at equal times, the right-hand side of Eq. (\ref{deltadelta}) vanishes by momentum conservation and the $1/q^2$ infrared divergence will not appear when calculating invariant quantities. 
For the three-point correlator, we obtain
\begin{eqnarray}
\Big< \delta_{\rm g}(\vq,\t)\delta_{\rm g}(\vk_1,\t_1)\delta_{\rm g}(\vk_2,\t_2)\Big>'_{q\to 0}
&=&- \Big< \delta_{\rm g}^L(\vq,\t) \delta_L(\vq,\t)\Big>' \left(\frac{D(\t_1)}{D(\tau)}
-\frac{D(\t_2)}{D(\tau)}
\right)\nonumber\\
&\times& \frac{{\vec q} \cdot \bk_1}{q^2}\Big<\delta_{\rm g}(\vk_1,\t_1)\delta_{\rm g}(\vk_2,\t_2)\Big>'.
\label{deltadeltadelta1}
\end{eqnarray}
Similarly, the dark matter correlators of the short wavelength modes in the background of the long wavelength mode perturbation should satisfy the relation 
\be
\Big<\delta(\t_1,\bx_1)\delta(\t_2,\bx_2)\cdots\delta(\t_n,\bx_n)\Big>_{v_L}=
\Big<\delta(\t_1',\bx_1')\delta(\t_2',\bx_2')\cdots\delta(\t_n',\bx_n')\Big>,
\ee
leading to \cite{KR,ppls,jorge}
\be
\Big< \delta(\vq,\t)\delta(\vk_1,\t_1)\cdots\delta(\vk_n,\t_n)\Big>'_{q\to 0}
= -P_{\delta_{\rm lin}}(q,\tau)\sum_{a=1}^n \frac{D(\t_a)}{D(\tau)} \frac{{\vec q} \cdot \bk_a}{q^2}\Big<\delta(\vk_1,\t_1)
\cdots\delta(\vk_n,t_n)\Big>',
\label{deltadelta2}
\ee
where 
$P_{\delta_{\rm lin}}(q,\tau)=(D(\tau)/D(\tau_{\rm in}))^2P_{\delta_{\rm lin}}(q,\tau_{\rm in})$ is the linear matter power spectrum. 
For the three-point correlator, we obtain
\be
\Big< \delta(\vq,\t)\delta(\vk_1,\t_1)\delta(\vk_2,\t_2)\Big>'_{q\to 0}
=- P_{\delta_{\rm lin}}(q,\tau) \left(\frac{D(\t_1)}{D(\tau)}-\frac{D(\t_2)}{D(\tau)}\right) \frac{{\vec q} \cdot \bk_1}{q^2}\Big<\delta(\vk_1,\t_1)
\delta(\vk_2,\t_2)\Big>'.
\label{deltadeltadeltam}
\ee 
Once more, we stress that these relations are valid at beyond linear order for the short wavelength modes 
which might well be in the non-perturbative regime.

\section{Consequences of the symmetries for the galaxy bias theory: non-local bias}
As the galaxy and dark matter overdensities equations of motion (\ref{fg1}-\ref{fg3}) and (\ref{fl1}-\ref{fl3}) are invariant 
under the set of transformations (\ref{gen}-\ref{gg3}), an immediate consequence is that one can construct scalar quantities, {\it i.e.} quantities $S(\vx,\t)$
which  upon the transformation (\ref{gen}) are such that
\be
S'(\vx,\t)-S(\vx,\t)=\vn\cdot\vec{\nabla} S(\vx,\t).
\ee
As the spatial gradients remain invariant, $\vec{\nabla}=\vec{\nabla}'$, 
one can easily realize that  there are the following scalar quantities in the dark matter sector at our disposal

\be
\fbox{$\displaystyle
\delta(\vx,\t),\,\, s_{ij}(\vx,\t)=\partial_i\partial_j\Phi(\vx,\t)-\frac{\delta_{ij}}{2}\Omega_{\rm m}{\cal H}^2\delta(\vx,\t),\,\,
t_{ij}(\vx,\t)=\partial_i v_j(\vx,\t)-\frac{\delta_{ij}}{3}\theta(\vx,\t)-\frac{2f}{3\Omega_{\rm m}\cal H}s_{ij}(\vx,\t)
$},
\nonumber\\
\label{invariants}
\ee
where $\theta(\vx,\t)=\vec{\nabla}\cdot\vv(\vx,\t)$, $f=\d \ln D/\d \ln a$ (with $D(a)$ is the growth factor as a function of the scale factor  $a$), we have removed the trace part from $\partial_i\partial_j\Phi(\vx,\t)$, which is nothing else than
the dark matter overdensity $\delta(\vx,\t)$, and $t_{ij}(\vx,\t)$ is vanishing at first-order in perturbation theory.
Notice that these quantities are scalars beyond the linear perturbation  theory as the symmetries identified in the previous section are valid 
at any order in perturbation theory. These symmetries are larger than the Galilean group identified in Ref. \cite{gal} for the
large-scale dynamics. Furthermore, upon constructing the invariant operators 
\be
D_\t^v=\frac{\partial}{\partial\t}+
\vv(\vx,\t)\cdot\vec{\nabla}\,\,\,\,{\rm and}\,\,\,\, 
D_\t^{v_{\rm g}}=\frac{\partial}{\partial\t}+\vv_{\rm g}(\vx,\t)\cdot\vec{\nabla},
\ee
one can construct two more scalar quantities 
\be
\vec{\nabla}\Phi(\vx,\t)+D_\t^v\vv(\vx,\t)+{\cal H}\vv(\vx,\t)\,\,\,\,{\rm and}\,\,\,\, 
\vec{\nabla}\Phi(\vx,\t)+D_\t^{v_{\rm g}}\vv_{\rm g}(\vx,\t)+{\cal H}\vv_{\rm g}(\vx,\t),
\ee
but they are nothing else than the momentum conservation quantities for the dark matter and the galaxy, respectively. They identically vanish on-shell and therefore are trivial. 

The set of invariants (\ref{invariants}) are useful in constructing a galaxy bias theory which goes beyond the local bias model \cite{local}. In the 
 latter the galaxy overdensity $\delta_{\rm g}(\vx,\t)$ 
 is  written as a completely general function $f[\delta(\vx,\t)]$ of the mass density perturbation $\delta(\vx,\t)$, and
then the function is Taylor expanded, with the unknown coefficients in the series becoming the bias parameters
 \be
 \delta_{\rm g}(\vx,\t)= f[\delta(\vx,\t)]=b_1(\t)\delta(\vx,\t)+\frac{b_2(\t)}{2}\delta^2(\vx,\t)+\cdots.
 \ee
 This local expansion, even though it is consistent with the  first invariant of the list (\ref{invariants}),
is expected to be valid   only on  very large scales and small times: as the symmetry dynamics allows the presence of more scalar quantities, 
there is no reason why they should not be generated along the subsequent evolution. This logic is the same 
 which applies in quantum field theory for operators: even though some of them are not present in the tree-level Lagrangian, they will appear at a certain order in perturbation theory unless they are forbidden by symmetry arguments. 
Therefore, assuming homogeneity and isotropy, one would expect
a more general bias model of the form (where the coefficients should be intended to be the renormalized ones \cite{MD})
 \begin{eqnarray}
\label{nonlocal}
 \delta_{\rm g}(\vx,\t)&=&b_1(\t)\delta(\vx,\t)+\frac{b_2(\t)}{2}\delta^2(\vx,\t)+c_{\nabla^2}(\tau)\nabla^2\delta(\vx,\t)+ c_{s^2}(\t)s_{ij}(\vx,\t)s^{ij}(\vx,\t)\nonumber\\
 &+&c_{s^2\nabla^2}\nabla^2(s_{ij}(\vx,\t)s^{ij}(\vx,\t))+c_{s^2\nabla^4}(\t)\nabla^2s_{ij}(\vx,\t)\nabla^2s^{ij}(\vx,\t)+
 \cdots,
 \end{eqnarray}
at quadratic order in the fields and the dots stand for the various other terms one can construct 
out of $s_{ij}(\vx,\t)$ and gradients.
 We see that an unavoidable  consequence of the symmetries of the problem is 
that  the bias model    is a non-local bias model \cite{des,MD,mat,nonlocalbias}; 
in fact the non-local expansion (\ref{nonlocal}) has been first proposed in Ref. \cite{MD} where 
the same invariants have been employed based on general arguments on the homogeneous gravitational field and dark matter velocity.
Some comments are in order:

\begin{itemize}
\item The series does not contain a piece proportional to
the gravitational potential $\Phi(\vx,\t)$: it is simply forbidden by the symmetries of the problem as $\Phi(\vx,\t)$ alone is not a scalar quantity.
\item The non-local bias expansion (\ref{nonlocal}) is not dictated solely by rotational invariance.
Instead it is the more generic symmetry (\ref{gen}) together with isotropy which fixes the form of the expansion.
\item The fluid equations during the matter-dominated period are also invariant under Lifshitz scalings of the form \cite{pd,sb,KR}
\begin{align}
 \t'=\lambda^z\t,&~~~\bx'=\lambda \bx,\label{ass1}\\
 \delta'(\bx,\t)&=\delta(\bx',\t'),\label{dss1}\\
 \delta_{\rm g}'(\bx,\t)&=\delta_{\rm g}(\bx',\t'),\label{dss2}\\
 \bv_{\rm g}'(\bx,\t)&=\lambda^{z-1}\bv(\bx',\t'), \label{vss1}\\
 \bv'(\bx,\t)&=\lambda^{z-1}\bv_{\rm g}(\bx',\t'), \label{vss2}\\
 \Phi'(\bx,\t)&=\lambda^{2(z-1)}\Phi(\bx',\t'), \label{gfss1}
 \end{align}
 for a generic Lifshitz weight $z$ and
 \be
\frac{\partial}{\partial \t}=\lambda^z\frac{\partial}{\partial \t'}, ~~~\bnabla=\lambda \bnabla'.
\ee
Therefore, the  Lifshitz weights of the bias coefficients should be
\be
[b_1]=[b_2]=0,\,\, [c_{\nabla^2}]=2,\,\, [c_{s^2}]=-4z,\,\,[c_{s^2\nabla^2}] = -2-4z,\,\, [c_{s^2\nabla^4}]=-4-4z.
\ee
These  Lifshitz weights fix the  time-behaviour of the corresponding coefficients for the growing mode. The fact that the  Lifshitz weights of $b_1$ and $b_2$ are vanishing tell us that their growing mode is constant in time. Indeed, it is well-known that at large times the system experiences the so-called debasing:  $b_1$  converges to unity and $b_2$ goes to zero.
Furthermore, the   Lifshitz weights fix   the corresponding time-behaviour of the remaining bias coefficients in their growing modes:  $c_{\nabla^2}$, 
$c_{s^2}$, $c_{s^2\nabla^2}$ and $c_{s^2\nabla^4}$
 should scale as $\tau^{2/z}$, $\tau^{4}$, $\tau^{(4z+2)/z}$ and $\tau^{(4z+4)/z}$, respectively. In particular, if one matches with the linear power spectrum of dark matter with spectral index $n$, one finds $z=4/(3+n)\simeq 1$ \cite{gal}. This explains why the non-local bias coefficients increase with time  during the matter-dominated period. Furthermore, if one expresses the non-local invariant
$s_{ij}(\vx,\t)s^{ij}(\vx,\t)$ at second-order in terms of the product of the linear overdensities, one finds that the Lifshitz symmetry imposes that 
the overall time scaling is $\tau^{-2}$ in a matter-dominated universe (once one goes to momentum space). This is precisely the scaling found in Ref.  \cite{mdu} and leads to the so-called debasing, that is at late times the bias converges to unity and matter and galaxy density fields agree.
\item As we already mentioned,  galaxies form at a range of redshifts and merge. So it would be interesting to 
extend our  results to the more realistic
case when the number density of galaxies
changes with redshift due to some arbitrary source including  the effects of galaxy formation
and merging. However, if the effective source is a function of the  scalar functions described above then our symmetry considerations
will apply to this more complete galaxy description too. For instance, in Ref. \cite{nonlocalbias} it was assumed that the effective
source was of the form $A(\t) j(\rho)$, where $A(\t)$ parametrizes the epoch of galaxy formation and $j(\rho)$ the effects of dark matter on galaxy formation and merging. In such a case the symmetry (\ref{gen}-\ref{gg3}) holds.
\item If the fluid equations are not invariant under the set of transformations (\ref{gen}-\ref{gg3}), as it happens for example in some modified theories of gravity to be discussed below, one expects other terms to appear in the bias expansion as the bias is scale-dependent. The possibility of testing the Poisson equation with a scale-dependent bias was discussed in \cite{hui-parfrey}.
\end{itemize}

\subsection{Consequences of the symmetries for the galaxy bias theory: independence from the smoothing  scale}
The galaxy consistency relation also holds  for smoothed quantities as the smoothing operation commutes with the coordinate transformation (\ref{gen}). Indeed, suppose we perform a smoothing operation with a window function around a sphere of radius $R_L$

\be
\delta_{R_L}(\vx)=\int\d^3y\,W\left(\left|\vy-\vx\right|,R_L\right)\,\delta(\vy),
\ee
where $W$ is the appropriate window function. Then  we have

\begin{eqnarray}
\label{demRL}
\delta_{R_L}(\vx')&=&\int\d^3y\,W\left(\left|\vy-\vx'\right|,R_L\right)\,\delta(\vy)=\int\d^3y'\,W\left(\left|\vy'-\vx'\right|,R_L\right)\,\delta(\vy')\nonumber\\
&=&\int\d^3y\,W\left(\left|\vy-\vx\right|,R_L\right)\,\delta'(\vy)\nonumber\\
&=&\delta'_{R_L}(\vx),
\end{eqnarray}
where in the last passage we have made use of the properties $\d^3y'=\d^3y$ and $(\vy'-\vx')=(\vy-\vx)$. This has an important consequence.
The local abundance of tracers (galaxies), at fixed proper time, is typically a function of the matter density field (and their spatial derivatives)
within a finite region of size $R_*\sim$ few Mpc for most tracers.  In the most models of bias, the overdensities of the tracers and dark matter
are understood as smoothed  on some scale large-scale $R_L$ so that they can
be interpreted as a counts-in-cells relation. However,  no additional
smoothing scale $R_L$ should enter in the final value of observables, {\it e.g.}   the correlation
functions on some scale $r$. This is because the smoothen scale $R_L$ is not physical, it is just a tool for the effective description and an arbitrary  ultra-violet cute-off \cite{fabian}.

The symmetries at our disposal provide a simple and straightforward way to show that the galaxy correlation functions do not depend
on the smoothing scale $R_L$.  Indeed, suppose we work in Fourier space and that we change the smoothing scale
$R_L$ by an infinitesimal amount $\delta R_L$. Correspondingly, the Fourier transformed window function will
be

\be
W\left[q(R_L+\delta\r)\right]\simeq W(q\r)+qW'(q R_L)\delta\r\simeq W(q\r)\, e^{qW'(q\r)/W(q\r)\delta\r},
\ee
where the prime stands for the differentiation with respect to the variable $q\r$. We can perform now an infinitesimal  coordinate transformation
$\vx'=\vx+\vn(\tau)$. According to the relation (\ref{demRL}), both tracers and dark matter overdensities will transform in momentum space as

\be
\delta'_{\vq,R_L}= \delta_{\vq,R_L}\,e^{i\vq\cdot\vn(\t)}=\delta_{\vq}W(q\r)\,e^{i\vq\cdot\vn(\t)}
\ee
and therefore

\be
\delta'_{\vq,R_L+\delta\r}= \delta_{\vq}W\left[q(\r+\delta\r)\right]\,e^{i\vq\cdot\vn(\t)}=
\delta_{\vq,R_L}\,e^{qW'(q\r)/W(q\r)\delta\r}\,e^{i\vq\cdot\vn(\t)}.
\ee
We see that if we choose the infinitesimal vector $\vn(\t)$ to be

\be
\vn(\t)=i\frac{\vq}{q}\frac{W'(q\r)}{W(q\r)}\delta\r,
\ee
we can compensate the infinitesimal change of the smoothing radius $\r$ and obtain that

\be
\delta'_{\vq,R_L+\delta\r}=\delta_{\vq,R_L}.
\ee
Since the correlators in the old and the new coordinate system have to be the same, we conclude that the dependence on the smoothing radius $\r$ drops off. Physically, this is due to the fact that changing the large-scale smoothing radius  by some amount amounts to include
(or exclude) more momentum modes into the smoothed overdensity. This addition (or subtraction) of momentum modes can be compensated by going to a coordinate system where these long wavelength modes have been removed (or added). This argument
holds in all epochs, included the $\Lambda$-dominated epoch. During the matter-dominated epoch we have another tool to
reach the same conclusion: the Lifshitz symmetry. Indeed, The change in the smoothing scale $\r$ can be compensated by a scaling transformation $\vx'=\lambda\vx$, or $\vq'=\vq/\lambda$. In such a case we have

\be
\delta'_{\vq,R_L+\delta\r}= \delta_{\vq/\lambda}W\left[q/\lambda(\r+\delta\r)\right].
\ee
If we choose $\lambda=\lambda_{\r}=(1+\delta\r/\r)$, we obtain

\be
\delta'_{\vq,R_L+\delta\r}= \delta_{\vq/\lambda_{\r},R_L},
\ee
and again we conclude that the smoothing scale dependence drops off when correlators are considered.

\subsection{Galaxy bispectrum consistency relation at tree-level}
Since the bias model (\ref{nonlocal}) respects the symmetries (\ref{gen}-\ref{gg3}), the three-point function of galaxies computed in this model should satisfy the consistency relation. In the next two subsections we explicitly verify that this is the case in perturbation theory at the tree and one-loop levels.
Let us start with the tree-level case.
The equal time DM-galaxy cross-correlation at second order in perturbation theory is
\begin{equation}
\pmgl{\vk,\t}{-\vk,\t} =b_1(\t) P_{\delta_{\rm lin}}(k,\t),
\label{puneq2}
\end{equation}
while the unequal time power spectrum is 
\begin{equation}
\pggl{\vk,\t_1}{-\vk,\t_2} = b_1(\t_1) b_1(\t_2)\pmml{\vk,\t_1}{-\vk,\t_2}.
\label{puneq2}
\end{equation}
The bispectrum of the galaxies at fourth order for unequal times is
\begin{eqnarray}
\left. \bggg{\vq,\t}{\vk_1,\t_1}{\vk_2,\t_2} \right|_{\delta^{(4)}}&=& b_1(\t_1) b_1(\t_2) \pmml{\vk_1,\t_1}{-\vk_1,\t} \pmml{\vk_2,\t_2}{-\vk_2,\t} \nonumber\\
& \times & \left[ 2 b_1(\t) F_S^{(2)}(\vk_1,\vk_2) + b_2(\t) + c_{s^2}(\t) S(\vk_1,\vk_2)\right] \nonumber\\
&+& \text{cyclic permutations of } (\t,\vq),(\t_1,\vk_1)\text{ and }(\t_2,\vk_2),
\end{eqnarray}
where 

\begin{eqnarray}
F_S^{(2)}(\vk_1,\vk_2)&=&\left[\frac{5}{7} +\frac{1}{2}(\vk_1\cdot\vk_2)\frac{k_1^2+k_2^2}{k_1^2 k_2^2}+\frac{2}{7}\frac{(\vk_1\cdot\vk_2)^2}{k_1^2 k_2^2}
\right],\nonumber\\
S(\vk_1,\vk_2)&=&-\frac{1}{3} +\frac{(\vk_1\cdot\vk_2)^2}{k_1^2 k_2^2}.
\end{eqnarray}
In the squeezed limit $q \rightarrow 0$, $\vec{k}_1 \simeq -\vec{k}_2$ we find
\begin{eqnarray}
\left. \bggg{\vq,\t}{\vk_1,\t_1}{\vk_2,\t_2}_{q\rightarrow 0} \right|_{\delta^{(4)}}&=& b_1(\t) b_1(\t_1) b_1(\t_2) \frac{\vq \cdot \vk_1}{q^2}\pmml{\vk_1,\t_1}{-\vk_1,\t_2}.\nonumber\\
& \times & \left(\pmml{\vq,\t}{-\vq,\t_2}-\pmml{\vq,\t}{-\vq,\t_1}\right) \\
& = & b_1(\t) b_1(\t_1) b_1(\t_2) \frac{\vec{q} \cdot \vk_1}{q^2} \pmml{\vk_1,\t_1}{-\vk_1,\t_2}\nonumber\\
& \times & P_{\delta_{\rm lin}}(q,\t) \left( \frac{D(\t_2)}{D(\t)}-\frac{D(\t_1)}{D(\t)}\right)\nonumber\\
&=&- \Big< \delta_{\rm g}^{(1)}(\vq,\t) \delta^{(1)}(\vq,\t)\Big>' \left(\frac{D(\t_1)}{D(\tau)}
-\frac{D(\t_2)}{D(\tau)}
\right)\nonumber\\
&\times& \frac{{\vec q} \cdot \bk_1}{q^2}\Big<\delta_{\rm g}^{(1)}(\vk_1,\t_1)\delta_{\rm g}^{(1)}(\vk_2,\t_2)\Big>'.
\end{eqnarray}
We observe that the consistency relation is trivially satisfied at linear order. One should note that non-local terms are sub-leading. We shall therefore ignore them in the one-loop computation and consider only the local-bias model in the following. 

\subsection{Galaxy bispectrum consistency relation at one-loop}
The check the consistency relation at one-loop, or more precisely at order 6 in perturbation theory, we have to evaluate the following expression
\begin{eqnarray}
\left. \bggg{\vq,\t}{\vk_1,\t_1}{\vk_2,\t_2}_{} \right|_{\delta^{(6)}} &=&\frac{\vec{q} \cdot \vec{k}_1}{q^2} \left( \frac{D(\tau_2)}{D(\tau)}-\frac{D(\tau_1)}{D(\tau)}\right) \nonumber \\
& \times & \left[\langle \delta^{(1)}(\vq,\t) \delta_{\rm g}(-\vq,\t) \rangle' \pgg{\vk_1,\t_1}{\vk_2,\t_2}\right]_{\delta^{(6)}}. \label{con6}\end{eqnarray}
We first consider the right-hand side where one should  be careful when expanding the square parenthesis. Indeed, even when $\delta$ is in the linear regime, $\delta_{\rm g}$ might be non-linear and higher order corrections to $\delta_{{\rm g}}$ have to be taken into account. The square parenthesis at order 4 in perturbation theory is therefore
\begin{eqnarray}
\left[\langle \delta^{(1)}(\vq,\t) \delta_{\rm g}(-\vq,\t) \rangle' \pgg{\vk_1,\t_1}{\vk_2,\t_2}\right]_{\delta^{(6)}}& = &  
\pmgl{\vq,\t}{-\vq,\t}\left. \pgg{\vk_1,\t_1}{\vk_2,\t_2}\right|_{\delta^{(4)}} 
\nonumber \\
&+ &
\langle \delta^{(1)}(\vq,\t) \delta_{\rm g}^{(3)}(-\vq,\t) \rangle'\pggl{\vk_1,\t_1}{\vk_2,\t_2},
\nonumber
\label{con}
\end{eqnarray}
where $ \delta_{\rm g}^{(3)}(\vq,\t) $ is the third order contribution to $ \delta_{\rm g}(\vq,\t)$.
The first term on the right-hand side can be written using the bias model  as
\begin{equation}
 \pmgl{\vq,\t}{-\vq,\t}\left. \pgg{\vk_1,\t_1}{\vk_2,\t_2}\right|_{\delta^{(4)}}  = b_1(\t) \pmml{\vq,\t}{-\vq,\t} \left( P_{{\rm g}}^{11} + P_{{\rm g}}^{12}+P_{{\rm g}}^{22} + P_{{\rm g}}^{13} \right) ,
\end{equation}
where
\begin{eqnarray}
P_{{\rm g}}^{11} &=& b_1(\t_1) b_1(\t_2) \left. \pmm{\vk_1,\t_1}{\vk_2,\t_2}\right|_{\delta^{(4)}}, \label{p11} \\
P_{{\rm g}}^{12} &=& \frac{1}{2} b_1(\t_1) b_2(\t_2) \int \d^3p \,  \left. \bmmm{\vk_1,\t_1}{\vec{p},\t_2}{\vk_2-\vec{p},\t_2} \right|_{\delta^{(4)}} \nonumber \\
&+& (\vk_1,\t_1)\leftrightarrow (\vk_2,\t_2),
\label{p12}\\
P_{{\rm g}}^{22} &=& \frac{b_2(\t_1)b_2(\t_2)}{2} \int \d^3p \, \pmml{\vec{p},\t_1}{-\vec{p},\t_2}\pmml{\vk_1-\vec{p},\t_1}{-\vk_1+\vec{p},\t_2},
\label{p22}\\
P_{{\rm g}}^{13} &=& \frac{b_1(\t_1) b_3(\t_2)}{2} \pmml{\vk_1,\t_1}{\vk_2,\t_2} \sigma_L^2(\t_2)\nonumber \\
&+& (\vk_1,\t_1)\leftrightarrow (\vk_2,\t_2), \label{p13}
\end{eqnarray}
while the second term is
\begin{eqnarray}
\langle \delta^{(1)}(\vq,\t) \delta_{\rm g}^{(3)}(-\vq,\t) \rangle'\pggl{\vk_1,\t_1}{\vk_2,\t_2} &=&  
\frac{1}{2}b_3(\t)b_1(\t_1)b_1(\t_2) \sigma_L^2(\t)\nonumber\\
&\times&\pmml{\vq,\t}{-\vq,\t}\pmml{\vk_1,\t_1}{\vk_2,\t_2},\nonumber\\
&&
\label{d3}
\end{eqnarray}
where we defined the linear variance
\begin{equation}
\sigma_L^2(\t) \equiv \int \d^3p\;P_{\delta_{\rm lin}}(p,\t).
\end{equation}
Let us now compute the left-hand side of  Eq. \eqref{con6} with the help the expressions one can find in Ref.  \cite{Sefusatti:2009qh} and check that the equality is satisfied. The unequal-time bispectrum $\bggg{\vq,\t}{\vk_1,\t_1}{\vk_2,\t_2}$ is composed by several terms which, for compactness, we will denote analogously to what done in Ref. \cite{Sefusatti:2009qh} by the notation
\begin{eqnarray}
\delta^D(\vq + \vk_1 + \vk_2)\bijk{ijk} &\equiv & \lim_\sq \left[ \frac{b_i(\t)b_j(\t_1)b_k(\t_2)}{i!j!k!}\langle \delta^{i}(\vq,\t) \delta^{j}(\vk_1,\t_1) \delta^{k}(\vk_2,\t_2)\rangle \right. \nonumber\\
&+&\left. \text{ permutations } (\t,\vq),(\t_1,\vk_1),(\t_2,\vk_2) \right].
\end{eqnarray}
In the following, we compute each term identifying the ones which behave at least $\mathcal{O}(q^{-1}P_{\delta_{\rm lin}}(q))$ as $\sq $.

$\bullet	$ The first term is
\begin{eqnarray}
\bijk{111} &=& b_1(\t)b_1(\t_1)b_1(\t_2) \left. \bmmm{\vq,\t}{\vk_1,\t_1}{\vk_2,\t_2}_{q \rightarrow 0} \right|_{\delta^{(6)}} \\
&=& b_1(\t) b_1(\t_1) b_1(\t_2) \left[ P_{\delta_{\rm lin}}(q,\t)  \frac{\vec{q} \cdot \vec{k}_1}{q^2} \left( \frac{D(\tau_2)}{D(\tau)}-\frac{D(\tau_1)}{D(\tau)}\right)\left. \pmm{\vk_1,\t_1}{\vk_2,\t_2}\right|_{\delta^{(4)}} \right], \nonumber
\label{111}
\end{eqnarray}
where we used the consistency relation for matter. This is exactly the term proportional to $P_{{\rm g}}^{11}$ in Eq. \eqref{p11} in the right-hand side of the consistency relation.

$\bullet	$ We express the trispectrum in the integral of the following term using the consistency relation
\begin{eqnarray}
\bijk{112,II} &=& \frac{1}{2} b_1(\t) b_1(\t_1)b_2(\t_2)  \int \d^3 p \, \left. \bmmmm{\vq, \t}{\vk_1,\t_1}{\vec{p},\t_2}{\vk_2-\vec{p},\t_2}_\sq \right|_{\delta^{(6)}} + \text{2 perm.} \nonumber\\
&=& \frac{1}{2} b_1(\t) b_1(\t_1)b_2(\t_2)  \int \d^3 p \, \left. \bmmmm{\vq, \t}{\vk_1,\t_1}{\vec{p},\t_2}{\vk_2-\vec{p},\t_2}_\sq \right|_{\delta^{(6)}} + (\t_1,\vk_1) \leftrightarrow (\t_2,\vk_2)\nonumber \\
&=& -P_{\delta_{\rm lin}}(q,\t) \int \d^3p \left[ \frac{\vq \cdot \vk_1}{q^2}\frac{D(\t_1)}{D(\t)} + \frac{\vq \cdot \vec{p}}{q^2}\frac{D(\t_2)}{D(\t)} +  \frac{\vq \cdot (\vk_2 - \vec{p})}{q^2}\frac{D(\t_2)}{D(\t)} \right]  \nonumber \\
&\times & \left. \bmmm{\vk_1, \t_1}{\vec{p},\t}{\vk_2-\vec{p},\t_2} \right|_{\delta^{(4)}} +  (\t_1,\vk_1) \leftrightarrow (\t_2,\vk_2)\nonumber \\
&=& -P_{\delta_{\rm lin}}(q,\t)  \frac{\vq \cdot \vk_1}{q^2}\left[ \frac{D(\t_1)}{D(\t)} - \frac{D(\t_2)}{D(\t)} \right]  \int \d^3p \left. \bmmm{\vk_1, \t_1}{\vec{p},\t}{\vk_2-\vec{p},\t_2} \right|_{\delta^{(4)}} \nonumber \\
&+ & (\t_1,\vk_1) \leftrightarrow (\t_2,\vk_2).
\label{112ii}
\end{eqnarray}
This is equal to the term proportional to $P_{{\rm g}}^{12}$ in Eq. \eqref{p12}. In the second line we ignored the permutation containing a bispectrum not in the squeezed limit.

$\bullet	$ The following contribution reproduces the term proportional to $P_{{\rm g}}^{22}$ in Eq. \eqref{p22}.
\begin{eqnarray}
\bijk{122,II} &=& b_1(\t) b_2(\t_1) b_2(\t_2) \int \d^3p \, \left.\bmmm{\vq,\t}{\vec{p},\t_1}{-\vq-\vec{p},\t_2}_\sq \right|_{\delta^{(4)}}  \nonumber \\
& \times & \pmml{\vk_1-\vec{p},\t_1}{-\vk_1+\vec{p},\t_2} + \text{ 2 perm.} \nonumber \\
&=& b_1(\t) b_2(\t_1)b_2(\t_2) \int \d^3p \, \left. \bmmm{\vq,\t}{\vec{p},\t_1}{\vq+\vec{p},\t_2}_\sq\right|_{\delta^{(4)}} \pmml{\vk_1-\vec{p},\t_1}{\vk_1-\vec{p},\t_2} \nonumber \\
&=&  -b_1(\t) b_2(\t_1)b_2(\t_2)P_{\delta_{\rm lin}}(q,\t)\left[ \frac{D(\t_1)}{D(\t)} - \frac{D(\t_2)}{D(\t)} \right] \nonumber\\
& \times & \int \d^3p \, \frac{\vq \cdot (\vk_1-\vec{p})}{q^2}\pmml{\vk_1-\vec{p},\t_1}{-\vk_1+\vec{p},\t_2}\pmml{\vec{p},\t_1}{-\vec{p},\t_2} \nonumber\\
&=& -\frac{1}{2} b_1(\t) b_2(\t_1)b_2(\t_2)P_{\delta_{\rm lin}}(q,\t)\left[ \frac{D(\t_1)}{D(\t)} - \frac{D(\t_2)}{D(\t)} \right]\frac{\vq \cdot \vk_1}{q^2}\nonumber\\
& \times & \int \d^3p \, \pmml{\vk_1-\vec{p},\t_1}{-\vk_1+\vec{p},\t_2}\pmml{\vec{p},\t_1}{-\vec{p},\t_2}.
\label{122ii}
\end{eqnarray}
In the second equality we kept the only permutation enhanced in the squeezed limit and in the third  we used the consistency relation for matter. Finally, we used the fact that
\begin{multline}
\int \d^3p \,\frac{\vq \cdot \vec{p}}{q^2} \pmml{\vk_1-\vec{p},\t_1}{-\vk_1+\vec{p},\t_2}\pmml{\vec{p},\t_1}{-\vec{p},\t_2} = \\
\frac{1}{2}\int \d^3p \,\frac{\vq \cdot \vk_1}{q^2} \pmml{\vk_1-\vec{p},\t_1}{-\vk_1+\vec{p},\t_2}\pmml{\vec{p},\t_1}{-\vec{p},\t_2},
\end{multline}
which can be deduced simply by doing the shift $\vec{p} \rightarrow \vk_1 - \vec{p}$.

$\bullet	$ The term below is enhanced in the squeezed limit as it contains a bispectrum at unequal times. It reproduces the term proportional to $P_{{\rm g}}^{13}$ in Eq. \eqref{p13} together with the term in Eq. \eqref{d3}
\begin{eqnarray}
\bijk{113,II}  &=& \left[ \frac{1}{2}b_1(\t)b_1(\t_1)b_3(\t_2) \sigma_L^2(\t_2) + \text{2 perm.} \right] \left. \bmmm{\vq,\t}{\vk_1,\t_1}{\vk_2,\t_2}_\sq \right|_{\delta^{(4)}} \nonumber \\
&=& \left[ \frac{1}{2}b_1(\t)b_1(\t_1)b_3(\t_2) \sigma_L^2(\t_2) + \text{2 perm.} \right] \nonumber \\
&\times & \frac{\vec{q} \cdot \vec{k}_1}{q^2} P_{\delta_{\rm lin}}(q,\t) \left( \frac{D(\t_2)}{D(\t)}-\frac{D(\t_1)}{D(\t)}\right) \pmml{\vk_1,\t_1}{\vk_2,\t_2}.
\label{113ii}
\end{eqnarray}

$\bullet	$ The term
\begin{eqnarray}
\bijk{112,I} = b_1^2 b_2 \left[ P(q) P(k_1) \right]_{\delta^{(6)}}+ \text{ 2 perm.} = \mathcal{O}(P_{\delta_{\rm lin}}(q))
\label{112i}
\end{eqnarray}
is not dominant because the $\mathcal{O}(\delta^{(4)})$ corrections to $P(q)$ are at most $\mathcal{O}(P_{\delta_{\rm lin}}(q))$ when $\sq$.

$\bullet	$ The following terms are not relevant because they involve either terms that are proportional to the non-squeezed bispectrum, which makes them at most $\mathcal{O}(P_{\delta_{\rm lin}}(q))$,  or terms containing the bispectrum in the squeezed limit at equal times, which vanish due to the consistency relation. $B$ denotes the bispectrum of matter
\begin{eqnarray}
\bijk{122,I}  &=& \frac{b_1 b_2^2}{2} P_{\delta_{\rm lin}}(k_1) \int \d^3p \, \left. B\right|_{\delta^{(4)}} (k_2,p,|\vk_2-\vec{p}|) + \text{ 5 perm.} = \mathcal{O}(P_{\delta_{\rm lin}}(q)),
\label{122i}\\
\bijk{113,I}  &=& \frac{b_1^2 b_3}{2} P_{\delta_{\rm lin}}(k_1) \int \d^3p \, \left. B\right|_{\delta^{(4)}}  (k_1,p,|\vk_1-\vec{p}|) + \text{ 5 perm.} = \mathcal{O}(P_{\delta_{\rm lin}}(q)).
\label{113i}
\end{eqnarray}

$\bullet	$ The following terms are not enhanced in the squeezed limit as they are just products of linear power spectra at this order
\begin{eqnarray}
\bijk{222}  &=& \frac{b_2^3}{2} \int \d^3p \, P_{\delta_{\rm lin}}(p)P_{\delta_{\rm lin}}(|\vq + \vec{p}|)P_{\delta_{\rm lin}}(|\vk_1-\vec{p}|) =  \mathcal{O}(1),
\label{222i}\\
\bijk{123,I}  &=& \frac{b_1 b_2 b_3}{2} P_{\delta_{\rm lin}}(q) \int \d^3p \, P_{\delta_{\rm lin}}(|\vk_1-\vec{p}|)P_{\delta_{\rm lin}}(p) + \text{ 5 perm.} = \mathcal{O}(P_{\delta_{\rm lin}}(q)) ,
\label{123i}\\
\bijk{123,II}  &=& b_1 b_2 b_3 P_{\delta_{\rm lin}}(q)P_{\delta_{\rm lin}}(k_1) \sigma_L^2 + \text{ 2 perm.} = \mathcal{O}(P_{\delta_{\rm lin}}(q)),
\label{123ii}\\
\bijk{114,I}  &=& \frac{b_1^2 b_4}{2} P_{\delta_{\rm lin}}(q)P_{\delta_{\rm lin}}(k_1)\sigma_L^2 + \text{ 2 perm.} = \mathcal{O}(P_{\delta_{\rm lin}}(q)).
\label{114i}
\end{eqnarray}

$\bullet	$ Finally, the two- and three-loops corrections ignored in Ref. \cite{Sefusatti:2009qh} are at most constant in the squeezed limit such that our result is fully correct at sixth order.
Overall, we conclude that the galaxy consistency relation is satisfied at tree- and one-loop level.

\section{Consistency relation of galaxy correlation functions in redshift space}
\noindent
 Let us discuss now discuss how the galaxy consistent relations are modified when going from real space to  redshift-space where experiments are performed.
The mapping from real-space position $\vx$ to redshift-space $\vs$ is given by \cite{scott}
\be
\vs=\vx+ \frac{1}{{\cal H}}(\vec{v}_{{\rm g}}\cdot \hat{x})\hat{x},
\ee
and the density field in redshift-space is obtained
by imposing mass conservation
\be
\label{mcc}
[1+\delta_{{\rm g}}(\vs)]\d^3 s=[1+\delta_{{\rm g}}(\vx)]\d^3 x.
\ee
In Fourier space the condition (\ref{mcc}) reads
\be
\delta_{\rm D}(\vk)+\delta_{{\rm g},s}({\vk})=\int\frac{\d^3 x}{(2\pi)^3}\,e^{-i \vk\cdot \vx}\,e^{-i\vec{v}_{{\rm g}}(\vx)\cdot \hat{x}\,(\vk\cdot\hat{x})/{\cal H}}[1+\delta_{{\rm g}}(\vx)].
\ee
By performing a spatial coordinate transformation $\vx\rightarrow \vx'=\vx+\vn(\tau)$ we know that, if $\delta_{{\rm g}}(\vx,\tau)$ and $\vv(\vx,\tau)$ satisfy the fluid equations, then 
$\delta_{{\rm g}}'(\vx,\tau)=\delta_{{\rm g}}(\vx',\tau)$ and $\vv'(\vx,\tau)=\vv(\vx',\tau)-\dot{\vn}$ do as well.
This implies that for the new solution we have
\begin{eqnarray}
\label{first}
\delta_{\rm D}(\vk)+\delta'_{{\rm g},s}(\vk)&=&\int\frac{\d^3 x}{(2\pi)^3}\,e^{-i \vk\cdot \vx}e^{-i\vec{v}_{{\rm g}}'(\vx)\cdot \hat{x}\,(\vk\cdot\hat{x})/{\cal H}}[1+\delta_{{\rm g}}'(\vx)]\nonumber\\
&=&\int\frac{\d^3 x}{(2\pi)^3}\,e^{-i \vk\cdot \vx}\,e^{-i\vec{v}_{{\rm g}}(\vx')\cdot \hat{x}\,(\vk\cdot\hat{x})/{\cal H}}\, e^{i\dot{\vec{n}}\cdot \hat{x}\,(\vk\cdot\hat{x})/{\cal H}}[1+\delta_{{\rm g}}(\vx')]\nonumber\\
&\simeq&\int\frac{\d^3 x}{(2\pi)^3}\,e^{-i \vk\cdot \vx}\,e^{-i\vec{v}_{{\rm g}}(\vx)\cdot \hat{x}\,(\vk\cdot\hat{x})/{\cal H}}\, e^{-i[(\vn\cdot\vec{\nabla})\vv_{{\rm g}}]\cdot\hx\,(\vk\cdot\hat{x})/{\cal H}}
e^{i\dot{\vec{n}}\cdot \hat{x}\,(\vk\cdot\hat{x})/{\cal H}}
[1+\delta_{{\rm g}}(\vx)+(\vn\cdot\vec{\nabla})\delta_{{\rm g}}(\vx)].\nonumber\\
&&
\end{eqnarray}
This expression is exact. Expanding for small $\vn(\t)$, we get
\begin{eqnarray}
\label{second}
\delta_{\rm D}(\vk)+\delta'_{{\rm g},s}(\vk)
&\simeq&\int\frac{\d^3 x}{(2\pi)^3}\,e^{-i \vk\cdot \vx}\,e^{-i\vec{v}_{{\rm g}}(\vx)\cdot \hat{x}\,(\vk\cdot\hat{x})/{\cal H}}
[1+\delta_{{\rm g}}(\vx)]\nonumber\\
&+&\frac{1}{{\cal H}}\int\frac{\d^3 x}{(2\pi)^3}\,e^{-i \vk\cdot \vx}\,e^{-i\vec{v}_{{\rm g}}(\vx)\cdot \hat{x}\,(\vk\cdot\hat{x})/{\cal H}}(\vk\cdot\hat{x})
\left\{-i\left[(\vn\cdot\vec{\nabla})\vv_{{\rm g}}(\vx)\right]\cdot \hx +i(\dot{\vec{n}}\cdot \hat{x})\right\}\,
[1+\delta_{{\rm g}}(\vx)]\nonumber\\
&+&\int\frac{\d^3 x}{(2\pi)^3}\,e^{-i \vk\cdot \vx}\,e^{-i\vec{v}_{{\rm g}}(\vx)\cdot \hat{x}\,(\vk\cdot\hat{x})/{\cal H}}\,(\vn\cdot\vec{\nabla})\delta_{{\rm g}}(\vx)\nonumber\\
&=&\delta_{\rm D}(\vk)+\delta_{{\rm g},s}(\vk)\nonumber\\
&+&\frac{1}{{\cal H}}\int\frac{\d^3 x}{(2\pi)^3}\,e^{-i \vk\cdot \vx}\,e^{-i\vec{v}_{{\rm g}}(\vx)\cdot \hat{x}\,(\vk\cdot\hat{x})/{\cal H}}(\vk\cdot\hat{x})
\left\{-i\left[(\vn\cdot\vec{\nabla})\vv_{{\rm g}}(\vx)\right]\cdot \hx +i(\dot{\vec{n}}\cdot \hat{x})\right\}\,
[1+\delta_{{\rm g}}(\vx)]\nonumber\\
&+&\int\frac{\d^3 x}{(2\pi)^3}\,e^{-i \vk\cdot \vx}\,e^{-i\vec{v}_{{\rm g}}(\vx)\cdot \hat{x}\,(\vk\cdot\hat{x})/{\cal H}}\,(\vn\cdot\vec{\nabla})\left[1+\delta_{{\rm g}}(\vx)\right].
\end{eqnarray}
If we start from this expression, upon integrating by parts  we find
\begin{eqnarray}
\label{aa}
\delta_{\rm D}(\vk)+\delta'_{{\rm g},s}(\vk)
&=&\delta_{\rm D}(\vk)+\delta_{{\rm g},s}(\vk)\nonumber\\
&+&\frac{i}{{\cal H}}\int\frac{\d^3 x}{(2\pi)^3}\,e^{-i \vk\cdot \vx}\,e^{-i\vec{v}_{{\rm g}}(\vx)\cdot \hat{x}\,(\vk\cdot\hat{x})/{\cal H}}(\vk\cdot\hat{x})
\left\{-\left[(\vn\cdot\vec{\nabla})\vv_{{\rm g}}(\vx)\right]\cdot \hx +(\dot{\vec{n}}\cdot \hat{x})\right\}\,
[1+\delta_{{\rm g}}(\vx)]\nonumber\\
&+&\frac{i}{{\cal H}}\int\frac{\d^3 x}{(2\pi)^3}\,e^{-i \vk\cdot \vx}\,e^{-i\vec{v}_{{\rm g}}(\vx)\cdot \hat{x}\,(\vk\cdot\hat{x})/{\cal H}}(\vn\cdot\vec{\nabla})
\left\{\vec{v}_{{\rm g}}(\vx)\cdot \hat{x}\,(\vk\cdot\hat{x})\right\}\,
[1+\delta_{{\rm g}}(\vx)]\nonumber\\
&+&i(\vk\cdot\vec{n})\,\delta_{{\rm g},s}(\vk).
\end{eqnarray}
This gives
\begin{eqnarray}
\label{aaa}
\delta_{\rm D}(\vk)+\delta'_{{\rm g},s}(\vk)
&=&\delta_{\rm D}(\vk)+\delta_{{\rm g},s}(\vk)+i(\vk\cdot\vec{n})\,\delta_{{\rm g},s}(\vk)\nonumber\\
&+&\frac{i}{{\cal H}}\int\frac{\d^3 x}{(2\pi)^3}\,e^{-i \vk\cdot \vx}\,e^{-i\vec{v}_{{\rm g}}(\vx)\cdot \hat{x}\,(\vk\cdot\hat{x})/{\cal H}}(\vk\cdot\hat{x})
\left\{\left[(\vn\cdot\vec{\nabla})\hx\right]\cdot \vv_{{\rm g}}(\vx) +(\dot{\vec{n}}\cdot \hat{x})\right\}\,
[1+\delta_{{\rm g}}(\vx)]\nonumber\\
&+&\frac{i}{{\cal H}}\int\frac{\d^3 x}{(2\pi)^3}\,e^{-i \vk\cdot \vx}\,e^{-i\vec{v}_{{\rm g}}(\vx)\cdot \hat{x}\,(\vk\cdot\hat{x})/{\cal H}}
[\vv_{{\rm g}}(\vx)\cdot\hx][1+\delta_{{\rm g}}(\vx)](\vn\cdot\vec{\nabla})(\vk\cdot\hx).
\end{eqnarray}
At this point we can use the  distant observer approximation, that is take the direction of the vector $\vx$ fixed, since it varies little from galaxy to galaxy: galaxies are relatively close to each other on the plane orthogonal to the line-of-sight. This amounts to taking $\bnabla \hat{x}\simeq \vec{0}$ and we finally obtain 

%
\begin{eqnarray}
\delta_{\rm D}(\vk)+\delta'_{{\rm g},s}(\vk)
&=&\delta_{\rm D}(\vk)+\delta_{{\rm g},s}(\vk)+i(\vk\cdot\vec{n})\,\delta_{{\rm g},s}(\vk)\nonumber\\
&+&\frac{i}{{\cal H}}\int\frac{\d^3 x}{(2\pi)^3}\,e^{-i \vk\cdot \vx}\,e^{-i\vec{v}_{{\rm g}}(\vx)\cdot \hat{x}\,(\vk\cdot\hat{x})/{\cal H}}(\vk\cdot\hat{x})
(\dot{\vec{n}}\cdot \hat{x})\,
[1+\delta_{{\rm g}}(\vx)].
\end{eqnarray}
Note that here the first line corresponds to the field transformation that gives rise to the consistency relation, which in redshift space will contain new terms induced by the second line of this expression. Using the explicit expression for $\dot{\vec{n}}$
$$
\dot{\vec{n}}(\t) = -\vec{v}_L(\t)  = -i \frac{\vq}{q^2} {\cal H} f(\t) \delta_L(\vq,\t),
$$
we obtain,
\begin{eqnarray}
\delta_{\rm D}(\vk)+\delta'_{{\rm g},s}(\vk)
&=&\delta_{\rm D}(\vk)+\delta_{{\rm g},s}(\vk)+\frac{\vk\cdot\vq}{q^2}\,\delta(\vq)\delta_{{\rm g},s}(\vk)\nonumber\\
&+&f\frac{k}{q}\delta(\vq)\int\frac{\d^3 x}{(2\pi)^3}\,e^{-i \vk\cdot \vx}\,e^{-i\vec{v}_{{\rm g}}(\vx)\cdot \hat{x}\,(\vk\cdot\hat{x})/{\cal H}}\mu_{\vk} \mu_{\vq} [1+\delta_{{\rm g}}(\vx)]\nonumber\\
&=&\delta_{\rm D}(\vk)+\delta_{{\rm g},s}(\vk)+\frac{\vk\cdot\vq}{q^2}\,\delta(\vq)\delta_{{\rm g},s}(\vk) + f\frac{k}{q}\mu_{\vk}\mu_{\vq} \delta(\vq)\delta_{{\rm g},s}(\vk)\,,
\end{eqnarray}
where $\mu_{\vk}$ is the cosine between the vector $\hat{k}$ and $\hat{x}$, and we used the distant observer approximation to take the cosines out of the integral in the second equality. We therefore obtain that in redshift space the consistency relations reads

\begin{equation}
\fbox{$\displaystyle
\frac{\Big< \delta_{{\rm g},s}(\vq,\t)\delta_{{\rm g},s}(\vk_1,\t_1)\cdots\delta_{{\rm g},s}(\vk_n,\t_n)\Big>'_{q\to 0}}{\Big< \delta_{{\rm g},s}(\vq,\t_1)\cdots\delta_{{\rm g},s}(\vk_n,\t_n)\Big>'}
=- \frac{P_{{\rm g},s}(q,\tau)}{b_1(\t)+f(\t)\mu_{\vq}^2}\sum_{a=1}^n\frac{D(\tau_a)}{D(\tau)}\left(\frac{\vq\cdot\vk_a}{q^2}+f(\tau_a)\frac{k_a}{q}\mu_{\vq}\mu_{\vk_a}\right)$},
\end{equation}
where we have used the linear relation \cite{sb}

\be
\delta_{{\rm g},s}(\vq,\t)=\left[b_1(\t)+f(\t)\mu_{\vq}^2\right]\delta(\vq,\tau).
\ee
In particular, the consistency relation for the bispectrum in redshift space explicitly reads
\begin{eqnarray}
\Big< \delta_{{\rm g},s}(\vq,\t)\delta_{{\rm g},s}(\vk_1,\t_1)\cdots \delta_{{\rm g},s}(\vk_2,\t_2)\Big>'_{q\to 0}
&=&- \frac{P_{{\rm g},s}(q,\tau)}{b_1(\t)+f(\t)\mu_{\vq}^2} \Big< \delta_{{\rm g},s}(\vk_1,\t_1)\delta_{{\rm g},s}(\vk_2,\t_2)\Big>' \bigg[\frac{{\vec q} \cdot \bk_1}{q^2} \left(\frac{D(\t_1)}{D(\tau)}-\frac{D(\t_2)}{D(\tau)}\right)\nonumber\\
&& +\,\left(\frac{D(\t_1)}{D(\tau)}f(\t_1) - \frac{D(\t_2)}{D(\tau)}f(\t_2)\right)\frac{k_1}{q}\mu_{\vk_1}\mu_{\vq}\bigg].
\label{deltadeltadeltam}
\end{eqnarray}
%


\section{Consequences of the symmetries  for the modified theories of gravity}
 Theories that (attempt to) explain the observed cosmic acceleration by modifying general relativity
all introduce a new scalar degree of freedom that is active on large scales, but is screened on small
scales to match experiments. All these theories introduce an extra light scalar field to modified gravity in the infrared. Typical examples
 are represented by the $f(R)$ theories \cite{fR}, which  are  equivalent to  classic
scalar-tensor theories \cite{chiba} and the screening effect takes place through the so-called chameleon mechanism \cite{ch},
 and by Galileon theories \cite{galileon} where the extra degree of freedom is appropriately
dressed through higher-derivative interactions  which decouple it form short-scale physics in accordance with solar system tests. 

 It has been recently realized that   in the modified gravity models where there is an efficient   screening phenomenon to make the set-up experimentally consistent there might also be   order unity violation of the EP \cite{HNS}. The galaxy and dark matter consistency relations
 are based on a coordinate transformation\footnote{Note that here the gradient of the long-wavelength mode $\vec{\nabla}\Phi_L$ is taken to be a constant vector in space, {\emph i.e.} recall that we are doing an expansion on the space variation of $\Phi_L$ and keep only terms linear in its gradients.} (in a matter-dominated period) \cite{KR,jorge}
 \be
  \t'=\t,&~~~\bx'=\bx+\int^\tau\d\eta\, \bv_L(\eta)=\bx+
    \frac{1}{6}\t^2\vec{\nabla}\Phi_L,
 \ee
 we are basically removing the time-dependent, but 
homogeneous gravitational force via a change of coordinates. This   corresponds to an homogeneous
acceleration transformation which allows to go to a free-falling observer, precisely the essence of the EP. Therefore, one expects
 a violation (or a spatial dependence)  of the galaxy consistency relation in modified gravity models where the screening mechanism is in action. 
 
Let us therefore  consider modifications of gravity that violate EP. As we said, in these models there exist  
 extra light  scalar fields which  effectively screen the scalar charge of objects as compared to unscreened objects of the same mass.
In other words, different objects of the same mass may have different scalar charge and they can move differently in the 
same environment violating the EP. The chameleon for example, has a potential such that 
it has long range forces outside of objects while it is massive in their interior. Therefore, the existence of such field
is consistent with solar system and fifth force tests but still can modify gravity at large distances. At any rate, 
we will assume here that there is such a screening mechanism and,  irrespectively of its origin, that it violates the EP.
The latter may be implemented  by modifying 
the energy-momentum conservation  as
\be
\nabla_\mu T^{\mu\nu}=f^\mu.
\ee
For the  non-relativistic dark matter fluid one therefore finds 
 \be
 &&\frac{\partial\rho(\vx,\t)}{\partial \t}+\bnabla\cdot[\rho(\vx,\t)\vec{v}(\vx,\t)]=0\, \label{e1}\\
&&\frac{\partial\vv(\vx,\t)}{\partial \t}+{\cal H}(\t)\vv(\vx,\t)+[\vec{v}(\vx,\t)\cdot\bnabla]\vec{v}(\vx,\t)=
-\bnabla \Phi(\vx,\t)
-\frac{q}{\rho}\bnabla \varphi(\vx,\t), \label{e2}
\ee
where we assumed that 
\be
f^i=-q\partial^i\varphi(\vx,\t),
\ee
with  $\varphi(\vx,\t)$  the scalar field that has environmental couplings 
that causes violation of the EP  and $q$ is the scalar charge density of the fluid. 
We  follow the parametrization introduced in Ref. \cite{HNS} and we  assume  $q=\epsilon \alpha \rho$, where $\alpha$  is a constant 
and $\epsilon$ is a parameter that describes the degree of screening ($\epsilon$ =0 for screened objects and $\epsilon=1$ 
for unscreened ones). 
We should supplement the above equations with the 
Poisson equation for the gravitational potential $\Phi(\vx,\t)$ and a corresponding equation for $\varphi(\vx,\t)$ which 
we write as
\be
 &&\nabla^2\Phi(\bx,\t)=4\pi Ga^2 \rho(\vx,\t),  \label{pois}\\
& & \nabla^2\varphi(\bx,\t)= \left(\frac{\partial V}{\partial\varphi}+8\pi G \alpha \rho(\vx,\t)\right) a^2,  \label{e3}
 \ee
 where $V(\varphi)$ is the scalar potential of the chameleon-like field. 
 We may now consider perturbations $\rho(\vx,\t)=\overline{\rho}(1+\delta(\vx,\t))$ and $\varphi(\vx,\t)=(\overline{\varphi}+\delta\varphi(\vx,\t))$ around the corresponding background 
 values $\overline{\rho}$ and $\overline{\varphi}$ and we find that these perturbations satisfy the equations 
 \be
 &&\frac{\partial\delta(\vx,\t)}{\partial \t}+\bnabla\cdot[(1+\delta(\vx,\t))\vec{v}(\vx,\t)]=0, \label{e01}\\
&&\frac{\partial\vv(\vx,\t)}{\partial \t}+{\cal H}(\t)\vv(\vx,\t)+[\vec{v}(\vx,\t)\cdot\bnabla]\vec{v}(\vx,\t)=-\bnabla \Phi(\vx,\t)-\epsilon\alpha\bnabla \delta\varphi(\vx,\t), \label{e02}\\
&&\nabla^2\Phi(\vx,\t)=4\pi Ga^2 \overline{\rho}\delta(\vx,\t),  \label{pois0}\\
& & \nabla^2\delta\varphi(\vx,\t)=\left(m^2\delta\varphi(\vx,\t)+8\pi G \alpha \overline{\rho}\delta(\vx,\t)\right) a^2,  \label{e03}
 \ee
 where 
 \be
 m^2(\bx,\t)=\left.\frac{\partial^2 V}{\partial\varphi^2}\right|_{\overline{\varphi}}
 \ee
 is the mass of the scalar field. 
 Restricting ourselves to the matter-dominated case, it can be checked that the Eqs. (\ref{e01}-\ref{e03}) 
 are invariant under the transformations
 \begin{align}
 \t'=\t,&~~~\bx'=\bx+\vec{n}(\t),\label{eq1}\\
 \delta'(\bx,\t)&=\delta(\bx',\t'),\label{eq2}\\
 \bv'(\bx,\t)&=\bv(\bx',\t')-\dot{\vec{n}}(\t), \label{eq3}\\
 \delta\varphi'(\bx,\t)&=\delta\varphi(\bx',\t'),\label{eq4}\\
\Phi'(\bx,\t)&=\Phi(\bx',\t')-\left(\ddot{\vec{n}}(\t)+{\cal H}(\t) \dot{\vec{n}}(\t)\right)\cdot \bx. \label{eq5}
 \end{align}
 As a result, it is  still possible to 
to remove a 
long wavelength mode for the velocity perturbation $\vv_L(\t,\vec{0})$ by properly choosing 
the vector ${\vec{n}}(\tau)$ in order. 
%
%
%
Indeed, in  the linear regime in momentum space the dynamical equations are given by 
 \be
 &&\frac{\partial \delta_L(\vq,\t)}{\partial\t}+i\vec{q}\cdot\vec{v}_L(\vq,\t)=0,\label{me1},\\
 && \frac{\partial\bv_L(\vq,\t)}{\partial\t}+{\cal{H}}(\t)\bv_L(\vq,\t)
 =-i\vec{q}\Big(\Phi_L(\vq,\t)+\alpha \epsilon\delta\varphi(\vq,\t)\Big),\label{me2}\\
 &&
q^2\Phi_L(\vq,\t)=-\frac{3}{2}{\cal H}^2\Omega_{\rm m}\delta_L(\vq,\t), \label{me3}\\
 &&  q^2\delta\varphi(\vq,\t)=-(m^2\delta\varphi)(\vq,\t)a^2-
3\alpha{\cal H}^2\Omega_{\rm m}\delta_L(\vq,\t), \label{me4}
 \ee
 where $(m^2\delta\varphi)(\vq,\t)$ is the Fourier mode of $m^2(\vx,\t)\delta\varphi(\vx,\t)$.
In particular consider the configurations shown in figure \ref{figure:long_wave}, in a region outside the spherical over-density of radius $R_0$  where the chameleon-like field is not 
screened and its mass may be neglected, one has
\begin{figure}
\centering
\includegraphics[width=0.65\textwidth]{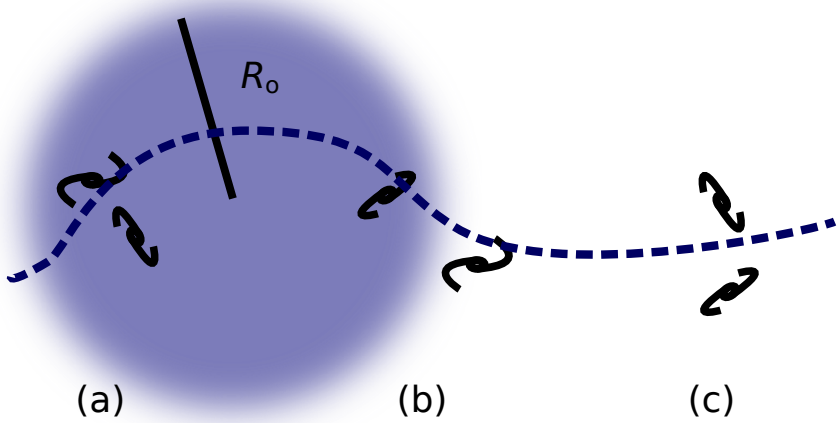}
\caption{\small Schematic representation of a large-scale spherical over-density of radius $R_0$ where the chameleon field is screened, and in the presence of a long-wavelength perturbation of the gravitational field (here represented by the dark blue dashed line). The consistency relation will be given by the correlation of the modulation of the power spectrum with the long-wavelength gravitational field. The case (a) corresponds to the case in which the galaxies are all in the screened region, Eq. (\ref{deltadelta22}), case (c) corresponds to the case in which all the galaxies are in the unscreened region, Eq. (\ref{delta-eq2}), and case (b) corresponds to the case in which there are both screened and unscreened galaxies, Eq. (\ref{delta-eq1}).}
\label{figure:long_wave}
\end{figure}
\be
\delta\varphi(\vq,\t)\simeq -
\frac{3}{q^2}{\cal H}^2\Omega_{\rm m}\delta_L(\vq,\t),
\ee
where the equation for the linear matter overdensity 
satisfies the equation\footnote{We use dots to denote derivatives with respect to conformal time.}
\be
 \ddot{\delta}_L+{\cal H}(\tau)\dot{\delta}_L-4\pi G a^2(1+2\alpha^2\epsilon)\overline{\rho}\delta_L=0, ~~~r\simgt R_0, 
 \label{qqq2}
 \ee
with solution  $\delta^>_L(\t)=D^>(\t)/D^>(\t_{\rm {in}})\delta(\t_{\rm in})$ where $D^>(\t)$ is 
the growth function for (\ref{qqq2}). 
%
On the contrary, in a screened region, where the 
 the field $\delta\varphi$ is  massive enough 
 so that $\varphi$ is not excited, but fixed 
 to some constant background value within a sphere of radius $R_0$. 
 In such a  case the equation for the overdensity is given by
 \be
 \ddot{\delta}_L+{\cal H}(\tau)\dot{\delta}_L-4\pi G a^2\overline{\rho}\delta_L=0, ~~~r\simlt R_0 \label{qq1}
 \ee
 and it  is solved by  $\delta^<_L(\t)=D^<(\t)/D^<(\t_{\rm in})\delta_L(\t_{\rm in})$.
Therefore $\vec{v}_L$ will be different  in the two regions  $r\simlt R_0$ and $r\simgt R_0$. As a result, two different vectors $\vec{n}$'s will be needed
 to generate (or remove) the long wave  velocity perturbation, one for $r\simlt R_0$ and the other $r\simgt R_0$: in the 
 presence of modified gravity exploiting the screening effect, it is not possible to find a spatially independent 
 vector $\vec{n}(\t)$ and the consistency relations must be violated for objects which are unscreened.
The vector  $\vec{n}(\t)$ is chosen such as to have  a free-falling frame, defined by 
\be
\ddot{\vec{n}}+{\cal H}\dot{\vec{n}}+ \vec{\nabla}\Phi=0.
\ee
The solution to this equation is 
\be
\dot{\vec{n}}(\t)=-i \frac{\vec{q}}{a(\t)}\int ^\t \d\eta \, a(\eta)\Phi(\vq,\eta)=
i \frac{\vec{q}}{q^2}\frac{1}{a(\t)}\int ^\t \d\eta \, a^3(\eta)4\pi G\bar{\rho}\delta_L(\vq,\eta).
\ee
Then, by using Eqs. (\ref{qqq2}) and (\ref{qq1}), we find that the free-falling frame is specified by 
\be
&&\vec{n}(\t)=i\frac{\vec{q}}{q^2}\delta^<_L(\vq,\t)\, , ~~~r\simlt R_0,\\
&&\vec{n}(\t)=i\frac{\vec{q}}{q^2}\frac{\delta^>_L(\vq,\t)}{1+2\alpha^2 \epsilon}\, , ~~~r \simgt R_0,
\ee
where we have indicated by $\delta^>_L$ and $\delta^<_L$ the dark matter overdensities in the two corresponding regions.
Consider for example $n$-galaxies within  a sphere of radius  $R\simgt R_0$ much smaller than the long 
wavelength mode of size $\sim 1/q$ and centered at the origin of the coordinates. 
Then,  if all points 
are at distances $r\simlt R_0$, then the consistency relation for the $n$-point correlator is
the one we already described
\be
\Big\langle \delta_{{\rm g}}(\vq,\t)\delta_{{\rm g}}(\vk_1,\t_1)\cdots\delta_{{\rm g}}(\vk_n,\t_n)\Big\rangle'_{q\to 0}
= -\sum_{a=1}^n \frac{{\vec q} \cdot \bk_a}{q^2} \Big< \delta^L_{\rm g}(\vq,\t)\delta^<_L(\vq,\t_a)\Big>\Big<\delta_{{\rm g}}(\vk_1,\t_1)
\cdots\delta_{{\rm g}}(\vk_n,\t_n)\Big>'\,.
\label{deltadelta22}
\ee 
If instead all points are at $r\simgt R_0$, we will have in this case 
\be
\Big\langle \delta_{{\rm g}}(\vq,\t)\delta_{{\rm g}}(\vk_1,\t_1)\cdots\delta_{{\rm g}}(\vk_n,\t_n)\Big\rangle'_{q\to 0}
= -\sum_{a=1}^n \frac{{\vec q} \cdot \bk_a}{q^2} \frac{\Big< \delta^L_{\rm g}(\vq,\t)\delta^>_L(\vq,\t_a)\Big>}{1+2\alpha^2\epsilon}\Big<\delta_{{\rm g}}(\vk_1,\t_1)
\cdots\delta_{{\rm g}}(\vk_n,\t_n)\Big>'\,.
\label{delta-eq2}
\ee
The case in which galaxies are both screened and unscreened is more complex\footnote{At the boundary between the over-dense region and the exterior waves of the scalar field will be generated and might propagate both to the interior and exterior. We will ignore these effects since we expect the scalar field to have small oscillations around the static solution deep inside the screened region, and in the unscreened region we expect the scalar field to go to a constant far from the boundary. Close to the boundary our results might not apply, but one can expect even larger violations to the consistency relation due to the gradient of the scalar field being large.}, however we expect a violation of the consistency relation due to the difference in the growth factor. Indeed, consider those configurations in which
$m$-galaxies are at $r\simgt R_0$ and $(n-m)$ are at $r\simlt R_0$, the consistency relation will be written as 
\begin{eqnarray}
\Big< \delta_{\rm g}(\vq,\t)\delta_{\rm g}(\vk_1,\t_1)\cdots\delta_{\rm g}(\vk_n,\t_n)\Big>'_{q\to 0}
&=& -
\!\left(\sum_{a=1}^{m} 
\frac{{\vec q} \cdot \bk_a}{q^2}\frac{\Big< \delta^L_{\rm g}(\vq,\t)\delta^>_L(\vq,\t_a)\Big>}{1+2\alpha^2 \epsilon}
\!+\!\sum_{a=m+1}^{n}
\frac{{\vec q} \cdot \bk_a}{q^2}\Big< \delta^L_{\rm g}(\vq,\t)\delta^<_L(\vq,\t_a)\Big>
\right)
\nonumber\\
&\times&\Big<\!\delta_{\rm g}(\vk_1,\t_1)
\!\cdots\!\delta_{\rm g}(\vk_n,\t_n)\!\Big>'.
\label{delta-eq1}
\end{eqnarray}
Notice that the right-hand side for the configuration (\ref{delta-eq1}) is not vanishing even for 
correlators at equal time for the $n$-points. This is due to the fact that 
the long wavelength  chameleon-like field correlates only with the overdensity located in the unscreened region, the one in the screened region being completely independent from the chameleon-like perturbation.
For instance, for $n=2$ and $m=1$, the corresponding bispectrum reads
\be
\fbox{$\displaystyle
\Big< \delta_{\rm g}(\vq,\t)\delta_{\rm g}(\vk_1,\t)\delta_{\rm g}(\vk_2,\t)\Big>'_{q\to 0}
=\left[ \frac{2\alpha^2\epsilon}{1+2\alpha^2 \epsilon}\, 
\Big< \delta^L_{\rm g}(\vq,\t)\overline{\delta}_L(\vq,\t)\Big>-\Big< \delta^L_{\rm g}(\vq,\t)\Delta{\delta}_L(\vq,\t)\Big> \right] \frac{{\vec q} \cdot \bk_1}{q^2}P_{\rm g}(\vk_1,\tau)$
},
\nonumber\\
&&
\label{delta-eq111}
\ee
where $\overline{\delta}_L=(\delta^<_L+\delta^>_L)/2$ and $\Delta\delta_L=(\delta^<_L-\delta^>_L)$. The latter is also suppressed by $\alpha^2\epsilon$ which 
%
therefore
gives an estimate of the violation of EP. 
Consider, for instance,   a cluster of galaxies of mass  $M\sim 10^{14.5\div 15}\,M_\odot$ and  radius $R_0\sim (2\div 10)$ Mpc. Inside it     $\Phi_{\rm cl}=-GM/R_0\sim -10^{-5}$ and one has  \cite{HNS}
 \be
 \frac{\overline{\varphi}}{2\alpha}\ll 10^{-6}\simlt   \frac{G M}{R_0},
 \ee
 where $\overline{\varphi}$ is the asymptotic background  value of the scalar $\varphi$  and  the upper bound comes from the solar system \cite{limit}. In such a dense object the scalar field is screened, 
 $\epsilon\simeq -\overline{\varphi}/(2\alpha\Phi_{\rm cl})\simlt 10^{-1}$, 
  and we may take $(n-m)$ galaxies residing there.   Away from the cluster there might be  small  $m$ galaxies with $\Phi_{\rm g}\sim -10^{-8}$ which are  unscreened (therefore preferably residing   in voids) and $\epsilon\simeq 1$. For this configuration, one expects to
 see a violation of the consistency relation as predicted by Eq. (\ref{delta-eq111}).
  Notice also that our considerations hold as long as  the Compton wavelength $m^{-1}$ associated to the 
  chamaleon-like field is larger than the scale where perturbations may be considered in the linear regime. At redshift $z=0$, there is a strong upper bound of about 1 Mpc on such Compton wavelength $m^{-1}(a_0)$ coming from the solar system tests 
 \cite{n1,n2}, implying that the  desired effects on the large
scale structure are restricted to non-linear scales. However, at higher redshifts a Compton wavelength of the form $m^{-1}(a)=m^{-1}(a_0)(a/a_0)^p$
 with $p<-3$ satisfies
the experimental constraints and can lead to a modified gravity regime on
large linear scales \cite{n1}. This  scaling of $m$ is  faster than the one deduced from the Lifshitz scaling of the scale $k_{\rm NL}(a)\sim a^{-2/(n+4)}$ (during matter-domination) at which cosmological perturbations become non-linear \cite{peebles,KR} and the condition
$m(a)<k_{\rm NL}(a)$ is easily attained going back in time. Notice also that the bound describes in Ref. \cite{n1,n2} does not hold in theories which screen by the Vainshtein mechanism, like the Galileon model, in which the scalar non-linearities
result from derivative interactions because the  screening condition
only holds up to  the  Vainshtein radius  \cite{n2}. In this class of theories though the violation of the EP for extended objects is tiny \cite{HNS}. 
So, an interesting question is how well one can measure a violation of the EP through the galaxy consistency relation. 
 Though an accurate  estimate  is beyond the scope of this paper, let us try to make a simple
 back-on-the-envelope computation by noting that the form of the bispectrum (\ref{delta-eq111}) is  almost the same one
  one obtains in the galaxy local bias model in the presence of a primordial local non-Gaussianity \cite{ks}  (see also Ref. \cite{ppls}).
Supposing that the combination $\alpha^2\epsilon$ is smaller than unity, one needs basically to identify (barring coefficient of order unity and assuming redshift $z=0$)
$\alpha^2\epsilon(\vq\cdot\vk_1)$ with $f_{\rm NL}H_0^2$, where $f_{\rm NL}$ is the non-linear coefficient parametrizing
the level of non-Gaussianity and $H_0$ is the present Hubble rate. The  Fisher matrix analysis applied to the galaxy (reduced) bispectrum  
performed in Ref. \cite{ks} has shown that one can measure $f_{\rm NL}$ up to ${\cal O}(10)$ for $k_1\sim k_{\rm max}\sim 0.1\, h$  Mpc$^{-1}$, being $k_{\rm max}$ the smallest scale scale included in the analysis. Therefore, again very roughly, we expect to be able to measure
deviation from the EP at redshifts $z\simgt 1$ of the order of ${\cal O}(10)(H_0/ \,k_{\rm max})\sim 10^{-3}(0.1\,h\,{\rm Mpc}^{-1}/k_{\rm max})$, where we have taken $q\sim 10^2\, H_0$.

Similar considerations apply also to more conventional modifications of gravity induced by scalars, like Brans-Dicke theory, or dilaton gravity. 
In these theories, in spite of the fact that there is a universal coupling of the scalar to matter, there is a violation
of the EP because different objects of the same mass may have different gravitational  binding energies.
However, this violation is subleading in the post-Newtonian approximation  for non-relativistic matter and it can only give 
order one effects in  strongly bound systems as binary systems and black holes \cite{Will}. 
To be more precise here, let us consider an   action of the general form 
\be
S=\int {\rm d}^4 x\sqrt{-g}\Big( f(R,\phi,X)+{\cal L}_{\rm m}\Big), \label{general}
\ee
where $f(R,\phi,X)$ is a function of the Ricci scale $R$, a scalar $\phi$ and its kinetic term  and
$X=-\frac{1}{2}\partial_\mu\phi\partial^\mu \phi$. This form of the action describes many models of modify gravity like Brans-Dicke
theory, dilaton gravity, $f(R)$ and many others. 
In this general class of models, the non-relativistic matter still satisfies Eqs.(\ref{fl1}-\ref{fl3}), where now 
$\Omega_{\rm m}=8\pi G_{\rm eff}\bar \rho a^2/3{\cal{H}}^2$ and  $G_{\rm eff}$ is an effective Newton constant which encodes
the modification of gravity  given by \cite{T}
 \be
 G_{\rm eff}(\t)=\frac{1}{8\pi F}
 \frac{f_{,X}+4\left(f_{,X}\frac{k^2}{a^2}\frac{F_{,R}}{F}+\frac{F^2_{,\phi}}{F}  \right) }
 {f_{,X}+3\left(f_{,X}\frac{k^2}{a^2}\frac{F_{,R}}{F}+\frac{F^2_{,\phi}}{F}  \right)}, ~~~F=\frac{\partial f}{\partial R}.
\ee
Therefore, when 
\be
f_{,X}\frac{k^2}{a^2}\frac{F_{,R}}{F}\ll 1,
\label{dd}
\ee
the effective Newton constant is only time dependent and it just modifies the temporal dependence of the local growth function
of the overdensity evolution. In this case, still, one may generate  a long wavelength velocity mode by a vector 
$\vec{n}(\t)$ as in
Eq. (\ref{long}). In the opposite case,   Eq. (\ref{dd}) is not satisfied and $G_{\rm eff}$ turns out to be space-dependent.  The overdensity 
$\delta$ turns out to be also space-dependent as well and there may be no $\vec{n}(\t)$ to generate a long wavelength velocity 
mode within  the sphere of radius $R_0$.  To see when this is possible, let us mention that there is a crossover scale 
when  the $k$-dependance of $G_{\rm eff}$ starts become strong and which is defined by
\be
R_0=\frac{a}{k}\approx \left(\frac{F,_R}{F} f,_X\right)^{1/2}.
\ee
If  $R\simlt R_0$, one may still define $\vec{n}(\t)$   and so long wavelength 
modes may be generated. On the other side if $R\simgt R_0$, {\it i.e.}  modification of gravity appears within the sphere, then 
there is no globally defined $\vec{n}(\t)$ inside the sphere of radius $R$, which will cause a modification of  the consistency relation. 
So the lesson here is that violation of the consistency relations is a signal of the  spatial dependence of the effective
Newton constant $G_{\rm eff}$ and of  a modification of gravity at large scales. 

We should also note that we have 
not considered here intrinsic violation of the EP, {\it i.e.} at the microscopic level \cite{FG1,Stubbs,Fayet,valeria}. One for example 
may consider the case of extra scalar, vector or tensor couplings to only one component, say   baryonic matter  or  dark matter. Such possibility has been considered recently in Ref. \cite{PP2} where it has been pointed out the interesting feature that if a large scale velocity bias exists between the different components new terms appear in the consistency relations with respect to the single species case. 
\section{Conclusions}
In this paper we have discussed the implications of the symmetry   enjoyed by the Newtonian equations of motion describing the 
 dark matter and galaxy fluids coupled through  gravity.  The fact that such symmetry applies to both galaxies and dark matter
is particularly welcome because one can reach conclusions which are independent from the galaxy bias. On the contrary, one can use the
power of the symmetry to deduce relevant informations on the theory of galaxy bias. In particular, we have shown that an unavoidable
consequence of the symmetries at our disposal  is that the bias is expected to be non-local. Furthermore, we have studied the modification (or violation) of the consistence relation in the case in which gravity is modified  because of the presence of extra degrees of freedom  propagating unscreened at large cosmological distances. Let us reiterate that our results are based on the assumption that
the galaxy  number is conserved. Eventually, one would like to extend our considerations by accounting for   phenomena like  halo formation and merging, nevertheless if the modification in the proper equations are such that the symmetries studied in this paper are preserved, {\it e.g.} if the new terms are a local function of the dark matter density, then our considerations remain valid. Also, apart from applying  to non-linear scales and directly to galaxies, our results have  the virtue of not being sensitive to the single stream approximation and to be valid also in the presence of velocity bias and/or vorticity (which is generated at higher-order in perturbation theory).
Therefore, assuming that primordial perturbations satisfy the consistency relations of \cite{creminelli2}, the observation of a deviation from the consistency relation for the bispectrum of galaxies, Eq. (\ref{deltadelta}), would signal either the inapplicability of the Eulerian bias model even including ``non-local'' terms as in Eq. (\ref{nonlocal}) or the violation of the EP in the underlying theory of gravity.

 \section*{Acknowledgments}
 When completing this work, Ref. \cite{PP2} appeared.  Our
results, when overlap is possible, agree with theirs. We thank M. Pietroni and M. Peloso for useful correspondence.
We acknowledge related work by P. Creminelli, J. Gleyzes, M. Simonovi\'c and F. Vernizzi and thank them for spotting an omission  in  the consistency relation in redshift
space in an earlier version of this draft.
H.P., J.N. and  A.R. are supported by the Swiss National
Science Foundation (SNSF), project `The non-Gaussian Universe" (project number: 200021140236). 
The  research of A.K. was implemented under the ``Aristeia" Action of the 
``Operational Programme Education and Lifelong Learning''
and is co-funded by the European 
Social Fund (ESF) and National Resources.  It is partially
supported by European Union's Seventh Framework Programme (FP7/2007-2013) under REA
grant agreement n. 329083.


\end{document}